\pdfoutput=1
\documentclass[prb,aps,twocolumn,amsmath,amssymb,floatfix,superscriptaddress]{revtex4}

\usepackage{graphicx}
\usepackage{xcolor}
\usepackage{soul}
\usepackage[colorlinks=true, citecolor=blue, urlcolor=blue]{hyperref}
\date{\today}

\begin{document}

%\captionsetup{justification=justified,singlelinecheck=false}

\title{Transport characteristics in Hermitian and non-Hermitian Fibonacci rings: A comparative study}
	
\author{Souvik Roy}

\email{souvikroy138@gmail.com}

\affiliation{School of Physical Sciences, National Institute of Science Education and Research Bhubaneswar, Jatni-752 050, India}

\affiliation{Homi Bhabha National Institute, Training School Complex, Anushaktinagar, Mumbai-400 094, India}
	
\author{Santanu K. Maiti}

\email{santanu.maiti@isical.ac.in}
	
\affiliation{Physics and Applied Mathematics Unit, Indian Statistical Institute, 203 Barrackpore Trunk Road, Kolkata-700 108, India}

\begin{abstract}
   We present an extensive theoretical analysis of transport and circular currents and the associated induced magnetic fields in Fibonacci rings, explored in both Hermitian and non-Hermitian descriptions, with particular attention to configurations preserving or breaking $\mathcal{PT}$ symmetry. By engineering physically balanced gain and loss following a Fibonacci sequence, we realize two distinct geometrical configurations in which the ring either preserve or explicitly break $\mathcal{PT}$ symmetry, and further explore complementary realizations obtained by reversing the signs of the on-site potentials. Using the nonequilibrium Green’s function (NEGF) formalism, we analyze transmission properties and bond current densities to quantify both transport and circulating currents. A comparison with the Hermitian limit establishes a clear baseline, where the ring supports only weak responses upon introducing disorder. In sharp contrast, non-Hermiticity leads to a pronounced amplification of transport and circular currents, and hence of the induced magnetic field. We further demonstrate that non-Hermitian transport is highly sensitive to gain and loss sign reversal and, in the non-$\mathcal{PT}$-symmetric case, exhibits an unconventional dependence on system size governed by the parity of the Fibonacci sequence and hopping correlations. Remarkably, the current does not decay monotonically with increasing system size, revealing a distinct scaling behavior absent in conventional Hermitian systems. Our results highlight non-Hermitian quasiperiodic rings as versatile platforms for engineering and amplifying current-driven magnetic responses through symmetry, topology, and gain-loss design.
\end{abstract}

\maketitle

\section{Introduction}
In recent times, the exploration of non-Hermitian (NH) frameworks endowed with $\mathcal{PT}$ symmetry~\cite{Ref1,Ref2,Ref3,Ref4,Ref5} has emerged as a cornerstone in understanding open quantum and classical systems, where environmental interactions play a decisive role. Non-Hermiticity typically originates from the inclusion of complex on-site potentials within the Hamiltonian, encapsulating the physical processes of gain and loss. When these counteracting effects are precisely balanced, the system retains its $\mathcal{PT}$ symmetry~\cite{Ref6,Ref7,Ref8,Ref9,Ref10}. This intriguing property has propelled extensive theoretical and experimental investigations across multiple platforms, most notably in photonic systems, where $\mathcal{PT}$-symmetric configurations have unveiled striking phenomena such as exceptional points, non-orthogonal eigenmodes, and unidirectional transparency. The vicinity of exceptional points further hosts a wealth of unconventional effects, including coherent diffusive transport, emergence of topological~\cite{Ref11,Ref12,Ref13,Ref14,Ref15} edge modes, chirality, and even the ability to arrest light propagation. Beyond optics, $\mathcal{PT}$-symmetric principles have profoundly influenced diverse areas ranging from atomic and molecular systems to electronic circuits and metamaterials, establishing a unifying paradigm for studying open-system dynamics~\cite{Ref16,Ref17,Ref18,Ref19,Ref20}. Despite these advancements, the intricate balance of gain and loss continues to fuel active research, as it not only governs spectral properties but also provides an exquisite control parameter for engineering novel quantum transport~\cite{Ref21,Ref22,Ref23,Ref24,Ref25} and wave manipulation phenomena.

In parallel, a growing body of research has focused on unraveling the intricate aspects of quantum transport in diverse tight-binding networks, highlighting the interplay between localization phenomena, transmission behavior, and correlated transport responses~\cite{Ref26,Ref27,Ref28,Ref29,Ref30}. Within these investigations, the incorporation of $\mathcal{PT}$-symmetric complex potentials has emerged as a pivotal factor capable of dramatically reshaping the underlying transport dynamics and coherence properties of such systems. Motivated by these findings, it becomes equally imperative to probe how environmental couplings, manifested through balanced gain and loss channels, modify the flow of charge carriers and influence current pathways across nanojunctions that integrate both elementary and topologically nontrivial loop geometries. Despite their conceptual and technological relevance, these aspects remain largely uncharted. Furthermore, analyzing the redistribution of currents and the emergence of bias-driven circulating currents in closed-loop architectures constitutes a crucial step toward a deeper understanding of mesoscopic quantum transport~\cite{Ref31,Ref32,Ref33,Ref34,Ref35,Ref36,Ref37,Ref38,Ref39,Ref40}, an endeavor that forms the central focus of our present investigation. In general, a system departs from Hermiticity once it is allowed to exchange energy or particles with its surrounding environment, leading to an effective open quantum description. Such non-Hermitian characteristics can be engineered by introducing site-dependent complex potentials, where the imaginary components correspond to physical processes of amplification (gain) and attenuation (loss)~\cite{svkadp,svksg,svkdavid}. An imbalance or directional preference in the hopping amplitudes may also give rise to a non-Hermitian scenario~\cite{Ref41,Ref42,Ref43,Ref44,Ref45,Ref46,Ref47,Ref48,Ref49,Ref50}. Within the optical and photonic domains, these mechanisms can be implemented in a controlled fashion through synthetic structures such as topolectrical or microwave resonator networks. The flexibility of such engineered systems allows the realization of balanced gain-loss configurations that exhibit $\mathcal{PT}$-symmetric phases and their associated symmetry-breaking transitions, thereby opening a versatile platform for exploring novel non-Hermitian transport phenomena~\cite{Ref51,Ref52,Ref53}.

Circular current established within a mesoscopic conducting ring is known to generate a substantial localized magnetic field, often reaching the order of a few Tesla, whose magnitude critically depends on the structural and electronic configuration~\cite{Ref54,Ref55} of the junction. Remarkably, this self-induced magnetic response offers a powerful mechanism to coherently manipulate a single spin positioned at or near the geometric center of the loop with atomic-scale accuracy, thereby presenting a promising route toward implementing spin-based qubits in quantum computational architectures. Although several theoretical and experimental efforts have investigated mechanisms for generating and modulating such circulating currents and the corresponding magnetic fields in various mesoscopic loop topologies~\cite{Ref56,Ref57,Ref58,Ref59,Ref60,Ref61,Ref62,Ref63,Ref64,Ref65,Ref66, Ref67, Ref68, Ref69, Ref70, Ref71, Ref72, Ref73}, the reported magnetic field strengths span a wide range, from a few millitesla up to multiple Tesla, depending on the underlying geometry and coupling parameters. 

Motivated by earlier studies, we develop a unified theoretical framework to investigate transport and circular currents, along with the associated induced magnetic fields, in Fibonacci quantum rings by systematically progressing from the Hermitian limit to the non-Hermitian regime. We begin by analyzing the Hermitian counterpart of the system in order to establish a clear reference point, against which the impact of non-Hermiticity can be unambiguously assessed. We then extend the model by introducing physically balanced gain and loss arranged according to a Fibonacci sequence, giving rise to non-Hermitian ring configurations that either preserve or explicitly break $\mathcal{PT}$ symmetry. Two distinct geometrical realizations are engineered by assigning different gain--loss patterns to the upper and lower arms of the ring, thereby allowing controlled access to $\mathcal{PT}$-symmetric and non-$\mathcal{PT}$-symmetric regimes. Within each geometry, we further construct complementary realizations by reversing the signs of the on-site potentials, enabling a systematic examination of configuration-dependent transport behavior. Using the nonequilibrium Green's function (NEGF) formalism, we compute transmission spectra and bond current densities to quantify both transport and circulating currents, with the latter generating an effective magnetic field threading the ring. The Hermitian analysis reveals that a symmetrically connected, perfectly ordered ring supports no circular current and only weak responses upon the introduction of disorder. This pronounced contrast with the non-Hermitian regime highlights the essential role of gain-loss engineering, system topology, and hopping asymmetry in achieving substantial amplification of transport and circular currents, and hence of the induced magnetic field, within an otherwise symmetric lead-ring-lead geometry.
\begin{figure}[ht]
%	\hskip 0.6in
\noindent
{\centering\resizebox*{8cm}{7cm}{\includegraphics{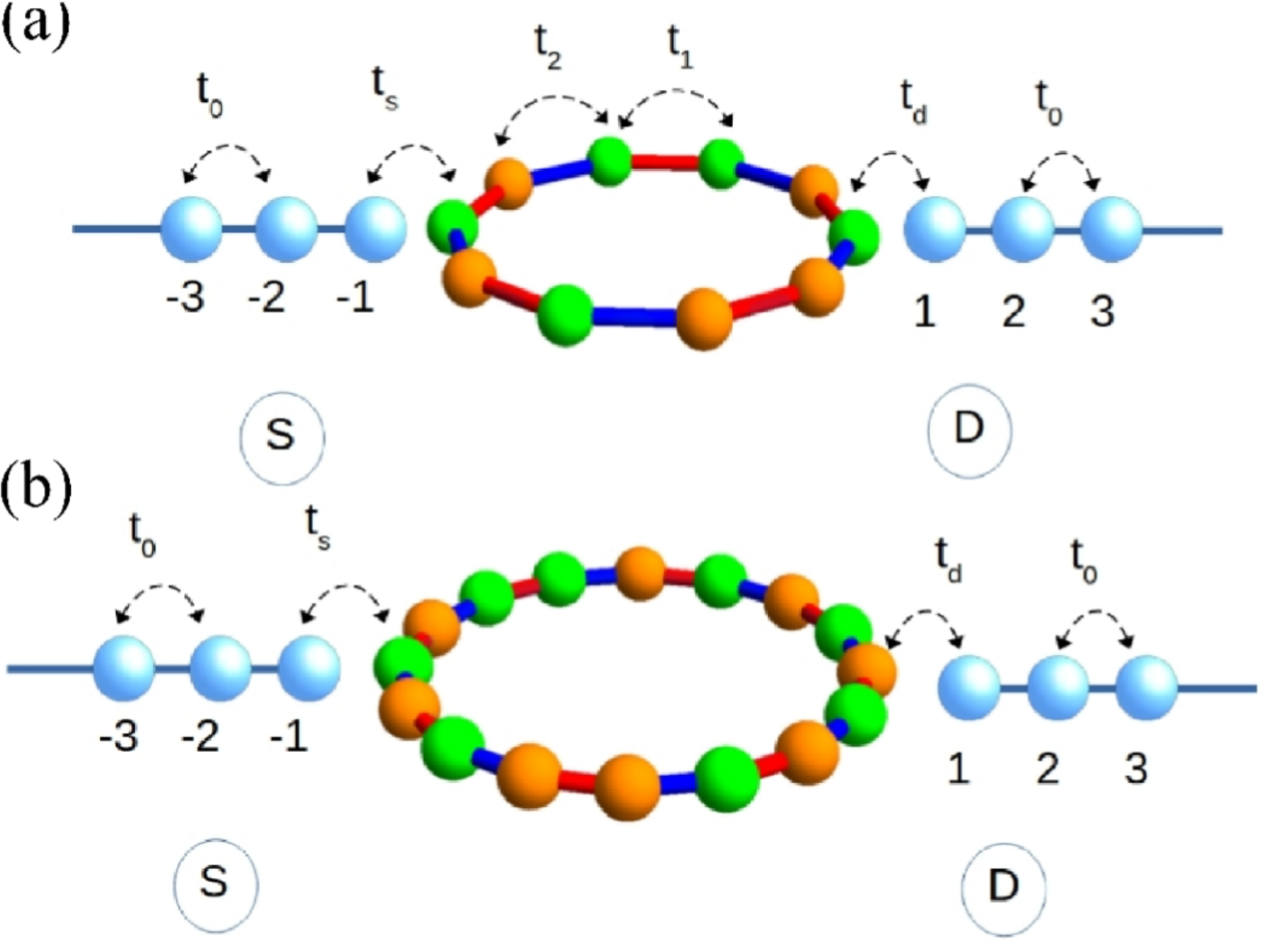}\par}}
\caption{(Color online). Schematic illustration of a quantum ring structure in which both the upper and lower arms are modulated by a Fibonacci quasiperiodic arrangement. Along the upper arm, the sites follow the sequence $\mathrm{A}\mathrm{B}\mathrm{A}\mathrm{A}\mathrm{B}\ldots$ as electrons propagate from the source to the drain, whereas the lower arm is characterized by the complementary sequence $\mathrm{B}\mathrm{A}\mathrm{B}\mathrm{B}\mathrm{A}\ldots$, again extending from the source to the drain. The two distinct lattice sites, $\mathrm{A}$ and $\mathrm{B}$, are represented by green and orange spheres, respectively. With the introduction of non-Hermitian contributions satisfying parity-time ($\mathcal{PT}$) symmetry, this configuration corresponds to a $\mathcal{PT}$-symmetric system. Subplot (a) corresponds to a system of size $L=10$, where each arm is characterized by the Fibonacci index $5$, whereas subplot (b) represents a larger system with $L=16$, with the Fibonacci index in each arm equal to $8$.}
\label{fig:sch}
\end{figure}
The main outcomes of our study include the following: (a) the transport current, circular current, and the associated induced magnetic field are markedly enhanced in the non-Hermitian regime compared to the Hermitian limit, highlighting the effectiveness of gain-loss engineering; (b) depending on the choice of system parameters, either the $\mathcal{PT}$-symmetric or the non-$\mathcal{PT}$-symmetric configuration can exhibit a stronger current response; (c) reversing the sign of the on-site potentials produces a significantly larger impact on transport in non-Hermitian systems than in their Hermitian counterparts, reflecting an increased sensitivity to local potential rearrangements; (d) the non-$\mathcal{PT}$-symmetric case exhibits a pronounced dependence on system size that stems from the parity (odd or even) of the underlying Fibonacci sequence, sign reversal of site energy and is further influenced by hopping correlations; and (e) in contrast to the commonly reported behavior in Hermitian systems, where currents typically decay monotonically with increasing size, the non-Hermitian Fibonacci rings studied here display a nonmonotonic size dependence of the current, revealing unconventional transport scaling driven by the interplay of Fibonacci ordering, topology, and non-Hermiticity.

% The remainder of the paper is organized as follows. In Sec.~I, we present the underlying motivation and clearly state the objectives of the present work. Section~II introduces the triangular ladder model and outlines the transport methodology based on the nonequilibrium Green’s function (NEGF) formalism. A comprehensive analysis and discussion of the obtained results are provided in Sec.~III, while Sec.~IV concludes the paper with a concise summary of the principal findings and their broader physical implications.

The remainder of the paper is organized as follows. Section~I outlines the motivation and objectives of the present work. In Sec.~II, we introduce the model and describe the transport formalism based on the NEGF approach. The results and their discussion are presented in Sec.~III, and Sec.~IV concludes the paper with a summary of the main findings and their physical implications.

\section{Model and theoretical formulation}
\begin{figure}[ht]
%	\hskip 0.6in
\noindent
{\centering\resizebox*{8cm}{7cm}{\includegraphics{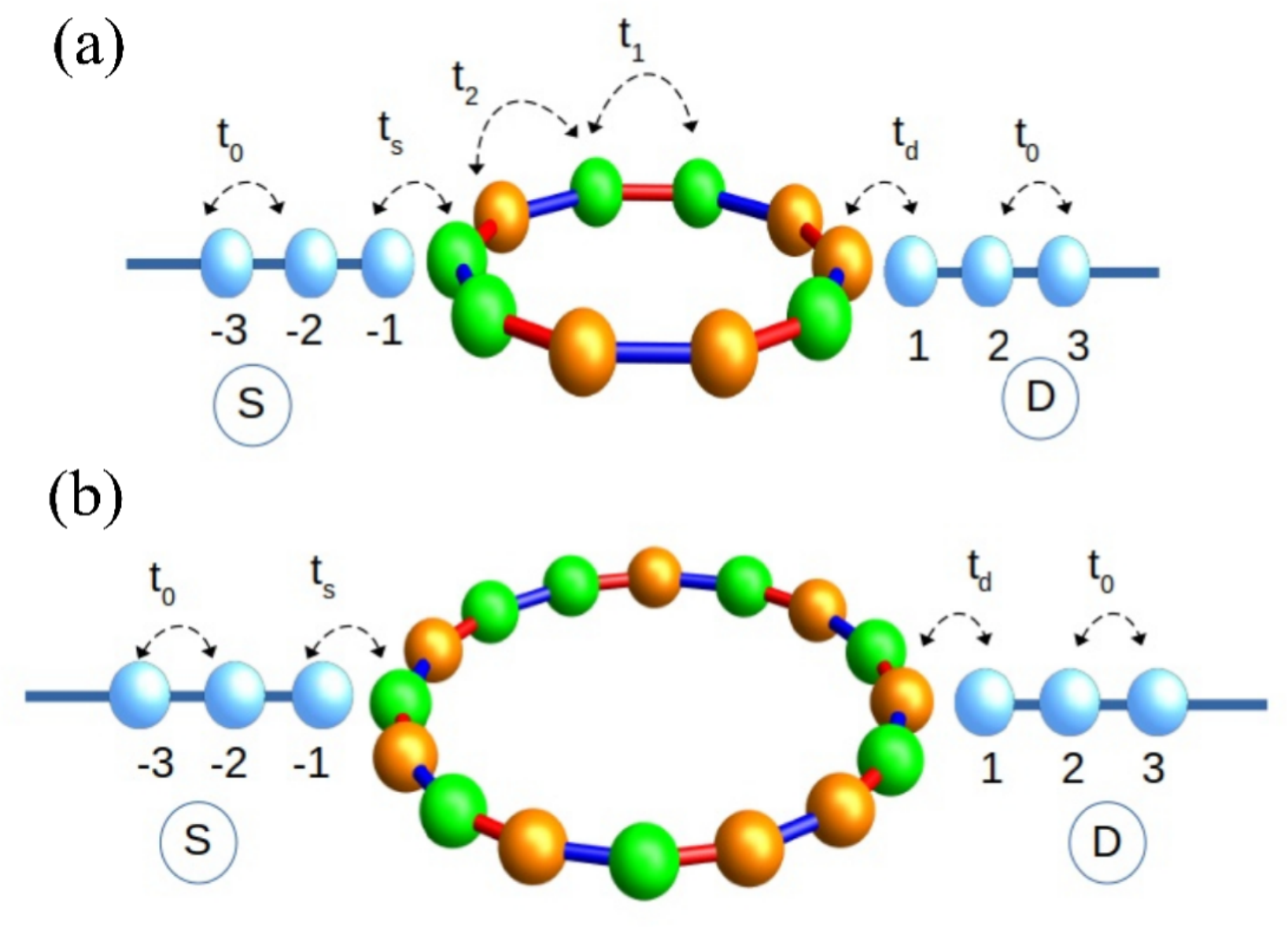}\par}}
\caption{(Color online). This schematic depicts a Fibonacci-modulated ring in which the lower arm is traversed in the reversed sequence, $\mathrm{B}\mathrm{A}\mathrm{B}\mathrm{B}\mathrm{A}\ldots$, from the drain to the source, while the upper arm retains the $\mathrm{A}\mathrm{B}\mathrm{A}\mathrm{A}\mathrm{B}\ldots$ ordering from source to drain. With the inclusion of non-Hermitian terms, this geometry realizes a non-$\mathcal{PT}$-symmetric configuration. The two subplots (a) and (b) correspond to $L=10$ and $L=16$, associated with Fibonacci indices $5$ and $8$ in each arm, respectively.
}
\label{fig:sch}
\end{figure}
In this section, we present a detailed description of the proposed model system, outlining the underlying tight-binding framework and the corresponding theoretical formalism employed to investigate its transport characteristics. We systematically develop the Hamiltonian of the system and introduce the methodological approach used to compute various transport properties, thereby establishing the foundation for the subsequent analysis and discussion. We consider two distinct physical scenarios, corresponding respectively to $\mathcal{PT}$-symmetric and non-$\mathcal{PT}$-symmetric configurations. For clarity, each scenario is illustrated using a distinct schematic representation. A detailed discussion of the underlying physical assumptions and the configuration engineering of the proposed model is presented below.

\noindent$\bullet$~\textbf{$\mathcal{PT}$-symmetric configuration:}  
In this configuration, the ring comprises two arms arranged in complementary Fibonacci sequences, with the upper arm traversed clockwise from source to drain and the lower arm traversed anticlockwise, thereby ensuring parity–time symmetry at the structural level. The site sequence along the upper arm follows
\[
\text{Upper arm: } A\,B\,A\,A\,B\,A\,\cdots,
\]
whereas the lower arm is its mirror-reversed counterpart,
\[
\text{Lower arm: } B\,A\,B\,B\,A\,B\,\cdots.
\]
% \begin{figure}[ht]
% %	\hskip 0.6in
% \noindent
% {\centering\resizebox*{8cm}{7cm}{\includegraphics{RINGSDF.pdf}\par}}
% \caption{(Color online). This schematic depicts a Fibonacci-modulated ring in which the lower arm is traversed in the reversed sequence, $\mathrm{B}\mathrm{A}\mathrm{B}\mathrm{B}\mathrm{A}\ldots$, from the drain to the source, while the upper arm retains the $\mathrm{A}\mathrm{B}\mathrm{A}\mathrm{A}\mathrm{B}\ldots$ ordering from source to drain. With the inclusion of non-Hermitian terms, this geometry realizes a non-$\mathcal{PT}$-symmetric configuration. The two subplots (a) and (b) correspond to $L=10$ and $L=16$, associated with Fibonacci indices $5$ and $8$ in each arm, respectively.
% }
% \label{fig:sch}
% \end{figure}
 % The upper arm obeys the order $A\,B\,A\,A\,B\,A\,\cdots$, while the lower arm follows the reverse arrangement $B\,A\,B\,B\,A\,B\,\cdots$, forming a mirror-symmetric configuration. 
% Each site of type $A$ and $B$ carries complex on-site potentials $+i\lambda$ and $-i\lambda$, respectively, introducing balanced gain and loss across the structure. 
A schematic illustration of the Fibonacci quantum ring considered in this work is presented in Fig.~1. The lattice sites of the ring are distributed along two distinct arms, each following a Fibonacci ordering, thereby endowing the system with an intrinsic structural aperiodicity. 
%In the non-Hermitian setting, we examine two representative realizations, one preserving $\mathcal{PT}$ symmetry and the other explicitly breaking it, whose geometric and gain–loss arrangements are detailed below.

\noindent$\bullet$~\textbf{Non-$\mathcal{PT}$-symmetric configuration:}  
In contrast, the non-$\mathcal{PT}$-symmetric geometry is obtained by arranging both arms of the ring to follow the Fibonacci sequence in the same (clockwise) sense, as shown in Fig.~2, thereby removing the counter-propagating traversal required for $\mathcal{PT}$ symmetry. Specifically, the upper arm is traversed from source to drain with the sequence $\mathrm{A}\mathrm{B}\mathrm{A}\mathrm{A}\mathrm{B}\mathrm{A}\ldots$, while the lower arm is traversed from drain to source with the complementary sequence $\mathrm{B}\mathrm{A}\mathrm{B}\mathrm{B}\mathrm{A}\mathrm{B}\ldots$.

% Here, electrons propagate clockwise along both the upper and lower arms from the source to the drain, leading to an explicit breaking of parity–time symmetry while retaining the underlying quasiperiodic order.
For each of the above configurations, we further define two complementary cases distinguished by the distribution of gain and loss. In \textit{case~1}, the gain–loss profile is assigned exactly as depicted in the schematic, with gain and loss placed on the corresponding lattice sites according to the prescribed Fibonacci sequence. In \textit{case~2}, this profile is completely inverted, such that every gain site in case~1 becomes a loss site and every loss site becomes a gain site. This deliberate interchange allows us to isolate and examine the sensitivity of transport and circular current characteristics to the sign reversal of the non-Hermitian on-site potentials, while keeping the underlying geometry and Fibonacci order unchanged.

The Fibonacci ring is connected symmetrically to two semi-infinite one-dimensional (1D) ideal electrodes, referred to as the source (S) and drain (D), as depicted in Fig.~1 and Fig.~2. 
Such symmetric coupling between the electrodes and the ring facilitates the emergence of bias-induced circulating currents within the non-Hermitian ring structure. 
The tight-binding (TB) formalism is employed to describe the entire setup, which is well-suited for capturing the essential physics of nanoscale systems with discrete lattice sites and Fibonacci order. The tight-binding Hamiltonian, along with the relevant theoretical framework, is detailed in the following subsections.

\subsection{Hamiltonian of the system}

The total Hamiltonian of the Fibonacci quantum ring connected to two semi-infinite electrodes is given by
\begin{equation}
\mathcal{H} = \mathcal{H}_{\mathrm{R}} + \mathcal{H}_{\mathrm{S}} + \mathcal{H}_{\mathrm{D}} + \mathcal{H}_{\mathrm{C}},
\label{eq:Htot}
\end{equation}
where \( \mathcal{H}_{\mathrm{R}} \) denotes the tight-binding Hamiltonian of the ring, \( \mathcal{H}_{\mathrm{S}} \) and \( \mathcal{H}_{\mathrm{D}} \) describe the source and drain electrodes, and \( \mathcal{H}_{\mathrm{C}} \) characterizes the coupling between the ring and the electrodes.

\subsubsection{Ring Hamiltonian}
% \noindent $\bullet$~\textbf{$\mathcal{PT}$-symmetric configuration:}

% The ring is constructed from two arms arranged according to a Fibonacci order. The upper half is traversed clockwise from the source to the drain, whereas the lower half is traversed anti-clockwise from the source to the drain. 
% The upper arm follows the sequence
% \[
% \text{Upper arm: } A\,B\,A\,A\,B\,A\,\cdots,
% \]
% while the lower arm exhibits its reversed configuration,
% \[
% \text{Lower arm: } B\,A\,B\,B\,A\,B\,\cdots.
% \]
% For case~1, the upper arm follows the above sequence, whereas for case~2, the sequences of the upper and lower arms are interchanged

% \noindent $\bullet$~\textbf{Non-$\mathcal{PT}$-symmetric configuration:}
% The ring is composed of two arms arranged according to a Fibonacci quasiperiodic sequence without enforcing parity--time ($\mathcal{PT}$) symmetry. Both the upper and lower arms are traversed clockwise from the source to the drain.
%, and the lower arm is not the spatially inverted counterpart of the upper one. The upper arm follows the sequence $A\,B\,A\,A\,B\,A\,\cdots$, while the lower arm is constructed independently with the same clockwise orientation, thereby explicitly breaking $\mathcal{PT}$ symmetry. For case~1, the Fibonacci sequence is assigned to the upper arm as specified above, with the lower arm following an identical ordering, whereas in case~2 the sequence assignments of the two arms are interchanged while preserving the clockwise traversal in both arms.

Each lattice site of the ring hosts one of two distinct on-site potentials, denoted by $\varepsilon_A$ and $\varepsilon_B$, corresponding to sites of type $A$ and $B$, respectively, as dictated by the underlying Fibonacci sequence. In the non-Hermitian regime, the on-site potentials are chosen to be purely imaginary and balanced in magnitude,
\begin{equation}
\varepsilon_A = + i \lambda, \qquad \varepsilon_B = - i \lambda,
\end{equation}
thereby introducing spatially distributed gain and loss of equal strength. In contrast, for the Hermitian limit the potentials are taken to be real with opposite signs,
\begin{equation}
\varepsilon_A = + \lambda, \qquad \varepsilon_B = - \lambda,
\end{equation}
which preserves Hermiticity while retaining the same Fibonacci modulation. The nearest-neighbor hopping amplitudes are characterized by two distinct parameters, $t_1$ and $t_2$, whose values depend on the local bond environment between adjacent sites, allowing for uniform or dimerized hopping configurations throughout the ring.

The tight-binding Hamiltonian of the ring takes the form
\begin{equation}
\mathcal{H}_{\mathrm{R}} =
\sum_{n} \varepsilon_{n}\, c^{\dagger}_{n} c_{n}
+ \sum_{\langle n,m \rangle} t_{nm}\,
\left( c^{\dagger}_{n} c_{m} + \mathrm{H.c.} \right),
\label{eq:HR}
\end{equation}
where \( \varepsilon_{n} = \varepsilon_{A} \) or \( \varepsilon_{B} \) depending on the type of site at position \( n \), and \( t_{nm} = t_{1} \) or \( t_{2} \) represents the corresponding hopping amplitude.  
Here, \( c^{\dagger}_{n} \) (\( c_{n} \)) denotes the creation (annihilation) operator of an electron at the \( n \)-th site.

% The upper and lower arms are designed such that they form mirror-symmetric Fibonacci chains, maintaining correlated but reversed site sequences.  
% The use of complex on-site energies with opposite imaginary parts introduces balanced gain and loss across the ring. 
\subsubsection{Electrode Hamiltonians and coupling}

The source and drain electrodes are modeled as ideal one-dimensional tight-binding leads:
\begin{equation}
\mathcal{H}_{\mathrm{S}} = \mathcal{H}_{\mathrm{D}} =
\varepsilon_{0}\sum_{n} d^{\dagger}_{n} d_{n}
+ t_{0} \sum_{\langle n,m\rangle}
\left( d^{\dagger}_{n} d_{m} + \mathrm{H.c.} \right),
\label{eq:HS_HD}
\end{equation}
where \( \varepsilon_{0} \) and \( t_{0} \) are the uniform on-site potential and nearest-neighbor hopping amplitude, respectively.  
The operators \( d^{\dagger}_{n} \) and \( d_{n} \) represent electron creation and annihilation operators at site \( n \) of the leads.

The coupling between the ring and the electrodes is described by
\begin{equation}
\mathcal{H}_{\mathrm{C}} =
\tau_{\mathrm{S}}\!\left( c^{\dagger}_{p} d_{0} + \mathrm{H.c.} \right)
+ \tau_{\mathrm{D}}\!\left( c^{\dagger}_{q} d_{N+1} + \mathrm{H.c.} \right),
\label{eq:HC}
\end{equation}
where \( \tau_{\mathrm{S}} \) and \( \tau_{\mathrm{D}} \) denote the tunneling amplitudes connecting the ring to the source and drain, respectively.  
The leads are attached to the ring at sites \( p \) and \( q \), which can be varied to investigate different contact geometries and quantum interference effects arising from the Fibonacci configuration.

\subsection{Transmission probability and junction current}

The electron transmission probability across the junction is determined within the Green’s function framework. It is expressed as~\cite{Ref59, Ref60}
\begin{equation}
\mathcal{T}(E) = \mathrm{Tr}\!\left[ \boldsymbol{\Gamma}_{\mathrm{S}}\, \mathcal{G}^{r}\, 
\boldsymbol{\Gamma}_{\mathrm{D}}\, \mathcal{G}^{a} \right],
\label{eq:T}
\end{equation}
where \( \boldsymbol{\Gamma}_{\mathrm{S}} \) and \( \boldsymbol{\Gamma}_{\mathrm{D}} \) denote the coupling matrices associated with the source and drain contacts, respectively. 
The quantities \( \mathcal{G}^{r} \) and \( \mathcal{G}^{a} = (\mathcal{G}^{r})^{\dagger} \) represent the retarded and advanced Green’s functions of the central scattering region. 
The retarded Green’s function is formulated as
\begin{equation}
\mathcal{G}^{r}(E) = \left[ (E + i0^{+})\mathbf{I} - \mathbf{H}_{\mathrm{R}} 
- \boldsymbol{\Sigma}_{\mathrm{S}} - \boldsymbol{\Sigma}_{\mathrm{D}} \right]^{-1},
\label{eq:Gr}
\end{equation}
where \( \mathbf{H}_{\mathrm{R}} \) denotes the Hamiltonian matrix of the device region, 
\( \boldsymbol{\Sigma}_{\mathrm{S}} \) and \( \boldsymbol{\Sigma}_{\mathrm{D}} \) are the self-energy corrections introduced by the semi-infinite source and drain electrodes, and \( E \) represents the energy of the incident electron.  

After computing the transmission function from Eq.~(\ref{eq:T}), the steady-state current through the nanojunction can be evaluated following the Landauer–Büttiker approach~\cite{Ref59, Ref60}. 
At zero temperature, the current flowing through the device under an applied bias voltage \( V \) is obtained as
\begin{equation}
I_{\mathrm{T}}(V) = \frac{2e}{h}
\int_{E_{\mathrm{F}} - eV/2}^{E_{\mathrm{F}} + eV/2}
\mathcal{T}(E)\, dE,
\label{eq:IT}
\end{equation}
where \( E_{\mathrm{F}} \) is the equilibrium Fermi level. 
The integration limits represent the bias window defined by the applied voltage.

\subsection{Circular current and magnetic response}

To evaluate the circular current in a closed-loop geometry, the bond currents between neighboring lattice sites must first be determined. 
The current passing through the bond connecting the sites \( i \) and \( j \) is expressed as~\cite{Ref34, Ref61}
\begin{equation}
I_{ij} = \frac{2e}{h}
\int_{E_{\mathrm{F}} - eV/2}^{E_{\mathrm{F}} + eV/2}
\mathcal{J}_{ij}(E)\, dE,
\label{eq:bond}
\end{equation}
where \( \mathcal{J}_{ij}(E) \) denotes the bond current density at energy \( E \), defined by
\begin{equation}
\mathcal{J}_{ij}(E) = \, \mathrm{Im}\!\left[ H_{ij}\, \mathcal{G}^{n}_{ji}(E) \right].
\label{eq:Jij}
\end{equation}
Here, \( H_{ij} \) represents the hopping element between the \( i \)-th and \( j \)-th sites in the Hamiltonian \( \mathbf{H}_{\mathrm{R}} \), and \( \mathcal{G}^{n} = \mathcal{G}^{r}\, \boldsymbol{\Gamma}_{\mathrm{S}}\, \mathcal{G}^{a} \) stands for the correlation (lesser) Green’s function of the system.

The net circular current \( I_{\mathrm{C}} \) circulating along the ring is determined by averaging over all \( N \) bonds constituting the loop:
\begin{equation}
I_{\mathrm{C}} = \frac{1}{N} \sum_{\langle i,j \rangle} I_{ij},
\label{eq:Ic}
\end{equation}
where the summation is performed over all nearest-neighbor pairs \( \langle i,j \rangle \).

The magnetic field generated by this circulating current at any arbitrary point \( \mathbf{r} \) inside the conducting ring can be evaluated using the Biot–Savart law~\cite{Ref63}:
\begin{equation}
\mathbf{B}(\mathbf{r}) =
\sum_{\langle i,j \rangle}
\frac{\mu_{0}}{4\pi}
\int
\frac{I_{ij}\, d\mathbf{r}' \times (\mathbf{r} - \mathbf{r}')}{|\mathbf{r} - \mathbf{r}'|^{3}},
\label{eq:BS}
\end{equation}
where \( \mu_{0} \) is the magnetic permeability of free space.  
This formulation directly relates the microscopic bond currents to the induced magnetic response within the ring structure.

\section{Results and Discussion}
In this section, we present a comprehensive analysis of the transport properties of both Hermitian and non-Hermitian Fibonacci rings, with particular emphasis on the transmission spectrum, junction (transport) current, and circulating (loop) current, along with the associated induced magnetic field. The transport current, circular current, and magnetic field are evaluated in units of $\mu\mathrm{A}$, $\mathrm{mA}$, and $\text{Tesla (T)}$, respectively. 
% To systematically assess the role of non-Hermiticity, we introduce two complementary configurations: in case~1 the on-site potential of type-$A$ sites is taken as $\epsilon_A=+i\lambda$ while that of type-$B$ sites is $\epsilon_B=-i\lambda$, whereas case~2 corresponds to an exact reversal of this gain--loss assignment. 
Throughout the analysis, the source and drain electrodes are coupled symmetrically to the ring, ensuring that any observed circular current and induced magnetic field arise purely from quantum interference, aperiodicity, and non-Hermitian effects rather than from contact asymmetry. For the asymmetric hopping case ($t_1 \neq t_2$), we fix $t_2 = 1$ and choose $t_1 = 1.5$ and $0.6$, whereas for the symmetric case ($t_1 = t_2$), both hopping amplitudes are set to unity. Energies are measured in units of eV. All remaining numerical parameters, model specifications, and unit conventions are introduced at the appropriate stages of the discussion to ensure clarity and continuity.

%\subsection{$\mathcal{PT}$-symmetric configuration}
\subsection{Hermitian Case}
We begin with the Hermitian limit and subsequently extend the analysis to the non-Hermitian regime in order to elucidate the role of non-Hermiticity. In the Hermitian case, we consider a quantum ring composed of two arms modulated by Fibonacci sequences, with the upper arm traversed from source to drain and the lower arm traversed from drain to source, thereby ensuring a controlled aperiodic modulation of the lattice sites. For \textit{case~1}, the on-site energies of the $\mathrm{A}$-type and $\mathrm{B}$-type sites are chosen as $\epsilon_A = +\lambda$ and $\epsilon_B = -\lambda$, respectively, while for \textit{case~2} the signs of the on-site potentials are interchanged.

\subsubsection{Transmission characteristics and junction current}
\vspace{0.15cm}
\begin{figure}[!htbp]
\centering
\includegraphics[width=8.8cm,height=8.2cm]{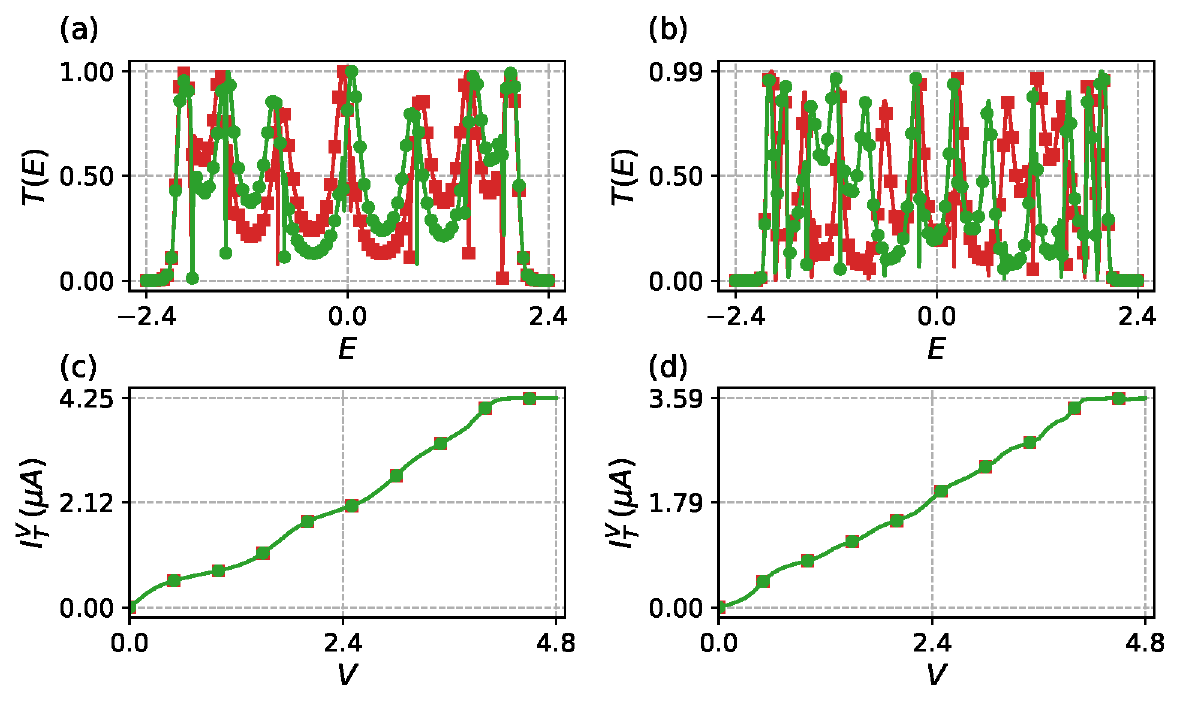}
\caption{(Color online).
Transmission coefficient and junction current characteristics for Fibonacci rings of different sizes. Panels~(a) and~(b) show the energy-dependent transmission spectra for system sizes $N=16$ and $N=26$, respectively for $\lambda=0.2$, where red and green curves represent case~1 and case~2 configurations.
Panels~(c) and~(d) display the corresponding junction current as a function of applied bias voltage.
}
\label{fig:PT_H_TRANS}
\end{figure}

The transport response of the Fibonacci ring is analyzed through the energy-dependent transmission spectra, as shown in Figs.~\ref{fig:PT_H_TRANS}(a) and~\ref{fig:PT_H_TRANS}(b) for system sizes $N=16$ and $N=26$, respectively. The red and green curves correspond to \textit{case~1} and \textit{case~2} configurations. 
%where the Fibonacci sequences in the upper and lower arms are either maintained or interchanged. 
Both cases exhibit nearly identical distributions of transmission resonances, implying that the Fibonacci backbone of the system predominantly dictates the transport behavior irrespective of the sequence orientation.

When a bias voltage is applied, the transport window broadens and additional transmission channels contribute to the total current. This behavior is evident in Figs.~\ref{fig:PT_H_TRANS}(c) and~\ref{fig:PT_H_TRANS}(d), where the current–voltage characteristics are presented for the same two system sizes. The junction current increases in a step-like fashion with voltage, each rise corresponding to the inclusion of new resonant states within the bias window. Remarkably, both cases yield almost indistinguishable current profiles across the entire voltage range, demonstrating that the overall transport is primarily governed by the global spectral symmetry arising from the Fibonacci geometry. This indicates a robust and stable conduction mechanism, resilient to sequence inversion or local structural rearrangements.

\subsubsection{Bond current density and circular current}
\vspace{0.15cm}
\begin{figure}[!htbp]
\centering
\includegraphics[width=8.8cm,height=8.0cm]{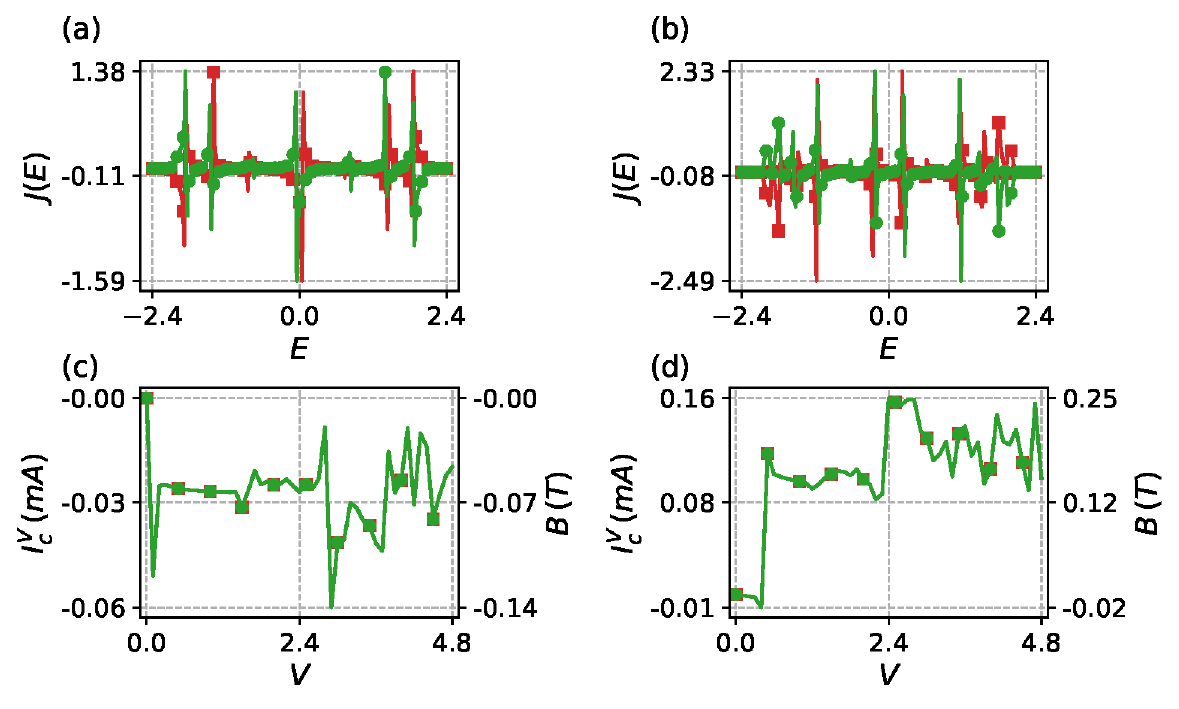}
\caption{(Color online).
Bond current density and circular current characteristics for Fibonacci rings.
Panels~(a) and~(b) show the energy-dependent bond current density for system sizes $N=16$ and $N=26$, respectively for $\lambda=0.2$, while panels~(c) and~(d) depict the circular current as a function of applied bias voltage. Red and green curves correspond to case~1 and case~2 configurations, respectively.
}
\label{fig:PT_H_BONDCUR}
\end{figure}

To gain a microscopic understanding of the internal current distribution, we analyze the bond current density and the corresponding circular current for the same system sizes considered in the previous plots, as shown in Figs.~\ref{fig:PT_H_BONDCUR}(a) and~\ref{fig:PT_H_BONDCUR}(b). The red and green curves illustrate the variation of the bond current density as a function of the incident electron energy $E$ for \textit{case~1} and \textit{case~2} configurations, respectively. The overall similarity between the two spectra highlights that the local current flow along the Fibonacci ring not affected much by the sequence inversion, reflecting the self-similar character of the Fibonacci order. Within certain energy windows, the two curves overlap significantly, corresponding to energy regimes dominated by extended and phase-coherent states.

When an external bias voltage is applied, the current density window expands, allowing additional localized states to contribute to charge transport and thereby altering the bond current density distribution. The corresponding circular current characteristics, shown in Figs.~\ref{fig:PT_H_BONDCUR}(c) and~\ref{fig:PT_H_BONDCUR}(d), clearly illustrate this behavior. As the applied voltage increases, subtle deviations appear between the two configurations, originating from delicate interference effects and phase modulations inherent to the Fibonacci configurations.
In conjunction with the circular current, we also present the induced magnetic field (plotted using a twin-$y$ axis) as a function of the applied voltage. The magnetic field attains a maximum value of approximately $0.15$~T for $N=16$ and $0.39$~T for $N=26$, reflecting the scaling of current circulation with system size.
Despite these quantitative differences, the overall current evolution remains qualitatively consistent for both configurations, reaffirming that the Fibonacci order collectively stabilizes the transport response even under sequence reversal.

\subsubsection{Variation of circular and transmission current with $\lambda$}
\vspace{0.15cm}
\begin{figure}[!htbp]
\centering
\includegraphics[width=8.8cm,height=8.0cm]{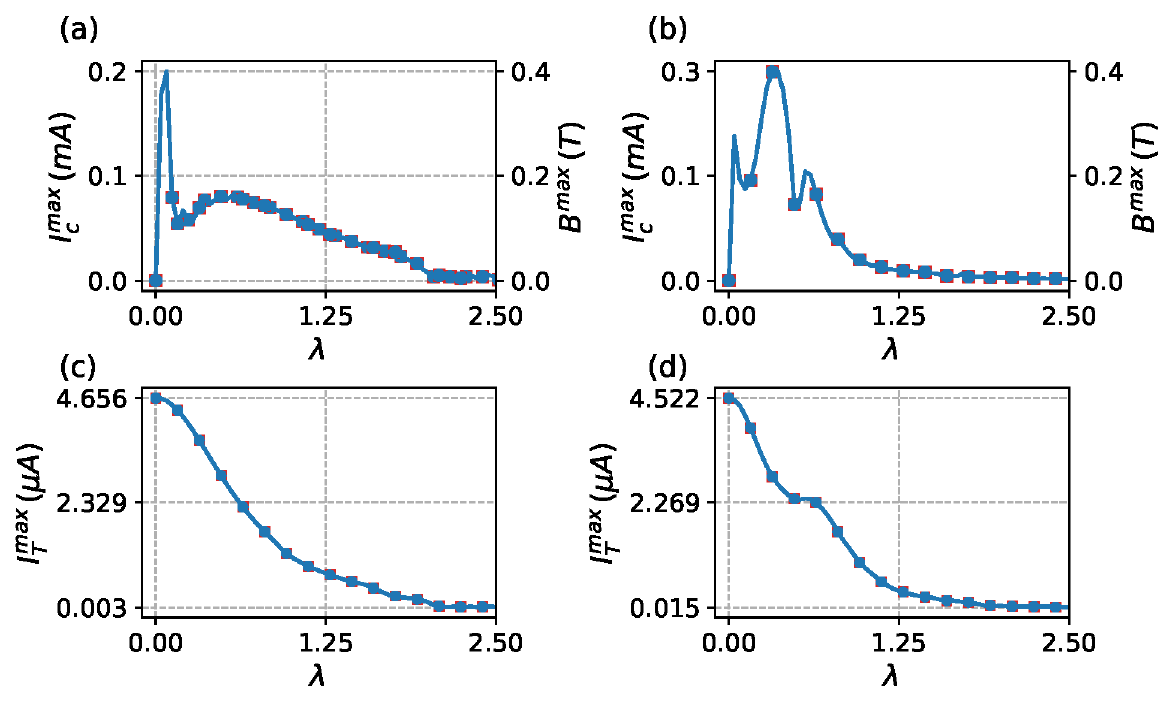}
\caption{(Color online).
Dependence of the maximum circular current and junction current on the potential strength $\lambda$ for Fibonacci rings of two different sizes ($N=16$ and $N=26$). 
Panels~(a) and~(b) show the variation of the maximum circular current with $\lambda$, while panels~(c) and~(d) present the corresponding junction currents. 
Red and blue curves denote case~1 and case~2 configurations, respectively, as discussed in the text. 
The induced magnetic field, derived from the circulating current, is plotted using twin-$y$ axes in panels~(a) and~(b), exhibiting peak values of approximately $0.5$~T and $0.4$~T for $N=16$ and $N=26$, respectively.}
\label{fig:PT_H_ICIT}
\end{figure}

To gain a deeper understanding of how Fibonacci modulation influences the transport dynamics, we analyze in Figs.~\ref{fig:PT_H_ICIT}(a) and~\ref{fig:PT_H_ICIT}(b) the variation of the maximum circular current with the strength of the potential $\lambda$ for two representative system sizes ($N=16$ and $N=26$). Similarly, Figs.~\ref{fig:PT_H_ICIT}(c) and~\ref{fig:PT_H_ICIT}(d) display the dependence of the maximum transport (or junction) current on $\lambda$ for the same systems. The red and blue curves correspond to \textit{case~1} and \textit{case~2} configurations, respectively, while the induced maximum magnetic field, obtained from the circulating current, is shown in the twin-$y$ axes of subplots~(a) and~(b). The maximum current in each case is determined by taking the largest value of the current over the entire applied voltage window. Notably, the induced magnetic field reaches values as high as $\sim 0.5~\mathrm{T}$ for $N=16$ and $\sim 0.4~\mathrm{T}$ for $N=26$, signifying a strong magnetoelectric coupling even in such framework.

A remarkable feature emerging from these plots is the close quantitative agreement between \textit{case~1} and \textit{case~2} across the full range of $\lambda$, underscoring that the transport properties are predominantly governed by the intrinsic Fibonacci topology rather than by the specific ordering of the Fibonacci sequences in the two arms. The circular current vanishes at $\lambda = 0$, which is physically intuitive: the system is perfectly symmetric in this limit, resulting in no net circulating current due to the absence of structural asymmetry and interference. As $\lambda$ increases, the circular current initially rises, reaching a maximum at an optimal disorder strength where constructive interference enhances current circulation, before declining again as stronger disorder drives the system toward localization.

In contrast, the transport or junction current exhibits the opposite trend. It is maximum at $\lambda = 0$, corresponding to the fully delocalized (perfectly ordered) limit, and progressively decreases as $\lambda$ increases. The suppression of current in the large-$\lambda$ regime reflects the transition from extended to localized electronic states, a hallmark of quasiperiodic systems where disorder-induced interference hinders coherent charge propagation. Overall, these results vividly demonstrate how $\lambda$ acts as a tunable control parameter that bridges delocalized and localized transport regimes, providing a coherent picture of current evolution and magnetic response in Fibonacci rings.

\noindent\textbf{Note:} We have additionally examined (results not shown) configurations in which both the upper and lower arms are traversed in the same direction, either clockwise or vice versa. The corresponding transport and circular current characteristics were found to be qualitatively similar to those presented above, differing only in the overall magnitude of the currents. Furthermore, we examined configurations with unequal hopping amplitudes ($t_1 \neq t_2$) for these traversal geometries. In these cases, no discernible qualitative distinction between \textit{case~1} and \textit{case~2} was observed. Instead, the transport response is dominated by the hopping asymmetry itself: the circular current attains its maximum value when $\lambda = 0$ and subsequently decreases in a nonlinear manner with increasing $\lambda$, reflecting the progressive suppression of coherent transport by disorder induced localization. In contrast, the transport current exhibits the same qualitative behavior as that observed in Figs.~\ref{fig:PT_H_ICIT}(c) and~\ref{fig:PT_H_ICIT}(d). In view of the absence of qualitatively new physical insights, these results are not presented here to maintain the compactness of the manuscript. 

Building upon the Hermitian analysis, we extend our investigation to the non-Hermitian regime by allowing the on-site potentials to acquire imaginary components, thereby introducing controlled gain and loss into the system. In this framework, transport phenomena are no longer governed solely by aperiodicity-induced localization; instead, they arise from a subtle competition between the intrinsic aperiodic order of the Fibonacci sequence, hopping correlations and non-Hermitian interference effects generated by complex onsite potentials. The hierarchical structure of the Fibonacci lattice plays a crucial role in modulating these interference processes, leading to qualitatively distinct transport responses compared to the Hermitian case. To capture these effects in a systematic manner, we analyze the system across different symmetry settings, starting from $\mathcal{PT}$-symmetric configurations and progressively moving toward explicitly non-$\mathcal{PT}$-symmetric regimes. The resulting transport characteristics, shaped by the interplay of symmetry, gain–loss imbalance, and Fibonacci order, are discussed in detail in the subsequent subsections.

\subsection{Non-Hermitian Case}
In the non-Hermitian regime, we consider two distinct cases, namely the $\mathcal{PT}$-symmetric and the non-$\mathcal{PT}$-symmetric configurations, whose geometrical realizations have already been introduced in the model section with the aid of schematic diagrams. In this regime, the real on-site potential $\lambda$ of the Hermitian limit is replaced by a purely imaginary term, $\pm \mathrm{i}\lambda$, representing balanced physical gain and loss, respectively. We first analyze the $\mathcal{PT}$-symmetric case and subsequently extend the discussion to the non-$\mathcal{PT}$-symmetric configuration, presenting the corresponding results in a systematic manner.
\subsubsection{$\mathcal{PT}$-symmetric configuration}
\begin{figure}[!htbp]
\centering
\includegraphics[width=8.8cm,height=8.0cm]{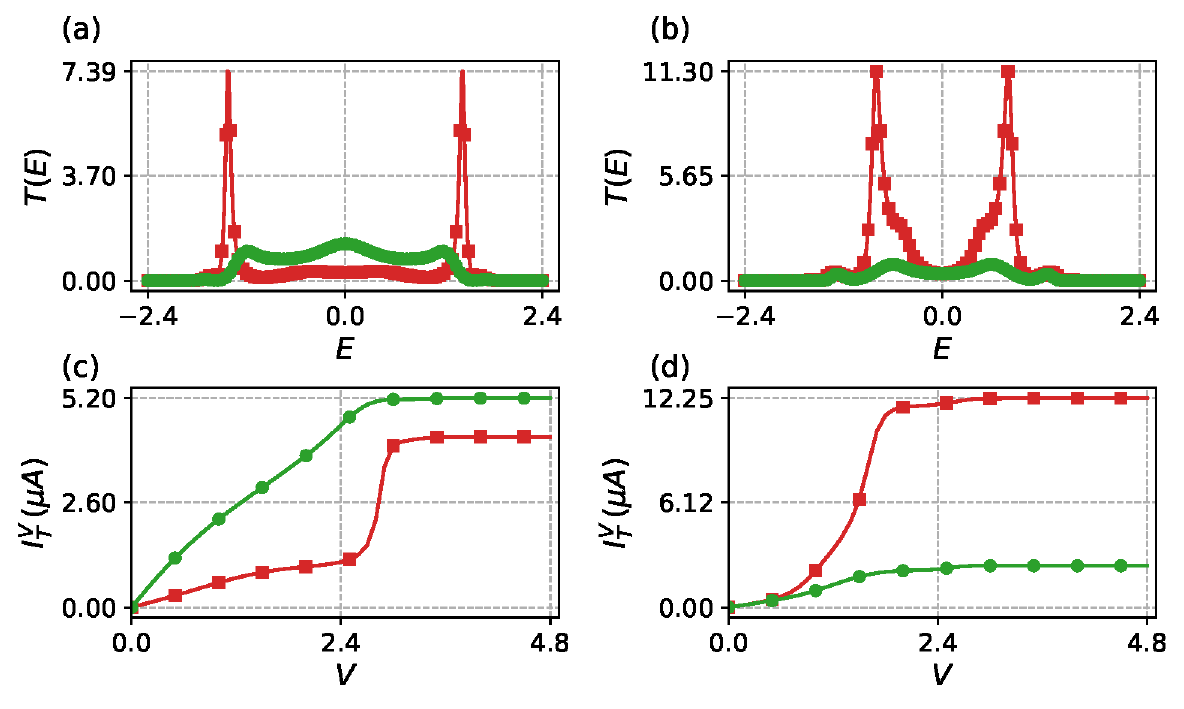}
\caption{(Color online).
Transmission spectra and current–voltage characteristics for the $\mathcal{PT}$-symmetric non-Hermitian lattice at $\lambda=1$. Panels (a) and (b) show the transmission probability as a function of incident energy for two distinct configurations, case~1 (red) and case~2 (blue), corresponding to system sizes $N=16$ and $N=26$, respectively. Panels (c) and (d) present the associated junction current obtained by integrating the transmission over the bias window.}
\label{fig:PT_NH_TRANS}
\end{figure}
In Figs.~\ref{fig:PT_NH_TRANS}(a)–(d), we depict the transmission spectra and the corresponding current–voltage characteristics for the $\mathcal{PT}$-symmetric non-Hermitian lattice under two representative configurations (\textit{case 1} and \textit{case 2}), represented by red and green curves for system sizes $N=16$ (left column) and $N=26$ (right column), respectively. The appearance of transmission amplitudes exceeding unity in Figs.~\ref{fig:PT_NH_TRANS}(a) and (b) is a direct manifestation of the non-Hermitian gain–loss interplay, which amplifies the outgoing wave components and gives rise to superunitary transmission, a characteristic feature of $\mathcal{PT}$-symmetric transport. For the smaller system ($N=16$), \textit{case 1} exhibits sharp resonant peaks confined to a narrow energy range, while \textit{case 2} displays a comparatively broader plateau with sustained transmission over an extended energy domain. As a result, the energy-integrated transport response or junction current, shown in Fig.~\ref{fig:PT_NH_TRANS}(c), becomes larger for \textit{case 2}, reflecting its enhanced spectral weight across the bias window. Interestingly, as the system size increases to $N=26$, the behavior undergoes a reversal: the transmission profile of \textit{case 1} now dominates across nearly maximum energy window, with higher resonance amplitudes than \textit{case 2}, leading to a larger net current in Fig.~\ref{fig:PT_NH_TRANS}(d). This size-dependent inversion of current hierarchy captures the intricate balance between non-Hermitian amplification, interference-induced delocalization, and spectral restructuring that collectively govern electron transport in $\mathcal{PT}$-symmetric non-Hermitian lattices. Another noteworthy observation is that, while the Hermitian model exhibits nearly identical behavior for \textit{case~1} and \textit{case~2}, the introduction of non-Hermiticity leads to a pronounced distinction between the two configurations, highlighting the crucial role of gain–loss imbalance in shaping the transport response.

\begin{figure}[!htbp]
\centering
\includegraphics[width=8.8cm,height=8.0cm]{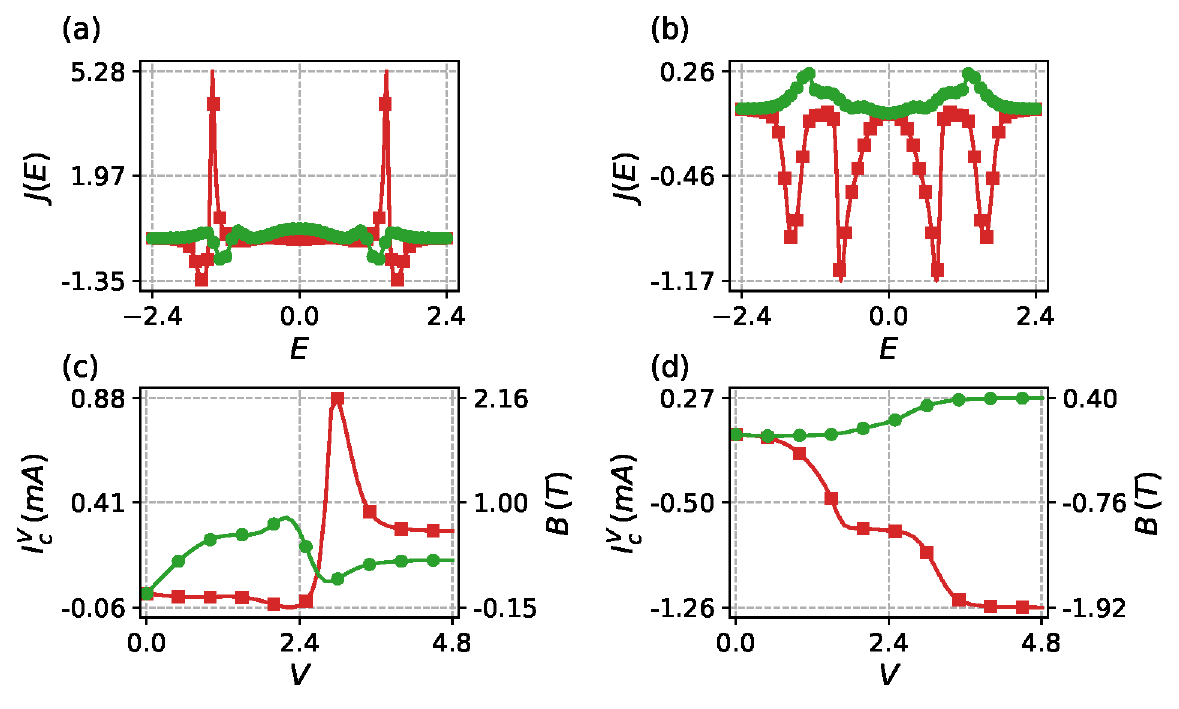}
\caption{(Color online).
Bond current density and circular current–voltage characteristics along with the induced magnetic field (shown on a twin $y$ axes) for the $\mathcal{PT}$-symmetric non-Hermitian lattice at $\lambda=1$. Panels~(a) and~(b) display the variation of bond current density with incident electron energy for two representative configurations—case~1 (red) and case~2 (blue)—corresponding to system sizes $N=16$ and $N=26$, respectively. Panels~(c) and~(d) illustrate the resulting circular current, obtained through energy integration over the applied bias window.}
\label{fig:PT_NH_BOND}
\end{figure}

In Fig.~\ref{fig:PT_NH_BOND}, we present a detailed analysis of the bond current density, circular current, and the corresponding induced magnetic field for the $\mathcal{PT}$-symmetric non-Hermitian lattice. Subplots~(a) and~(b) depict the variation of bond current density with incident electron energy, while subplots~(c) and~(d) illustrate the circular current and the associated magnetic field (shown on a twin $y$ axes) as functions of the applied bias voltage. The left and right columns correspond to system sizes $N=16$ and $N=26$, respectively, with \textit{case~1} and \textit{case~2} configurations represented by red and green curves. For $N=16$, the bond current density for \textit{case~2} exhibits a pronounced dominance near the zero-energy region, as evident in subplot~(a). Consequently, the circular current of \textit{case~2} surpasses that of \textit{case~1} up to a bias voltage of approximately $2.4$~V, as shown in subplot~(c). Beyond this range, the current behavior reverses, with \textit{case~1} taking the lead, a trend consistent with the underlying bond current spectra. In contrast, for the larger system ($N=26$), \textit{case~1} consistently exhibits stronger bond current density as well as circular current and induced magnetic field throughout the investigated parameter space, as seen in subplots~(b) and~(d). An additional noteworthy observation is that the bond current density, and hence the direction of the circular current, changes sign between the two configurations: \textit{case~1} yields a negative current, while \textit{case~2} produces a positive one. Although the sign merely reflects the direction of current circulation, this reversal signifies a controllable switching of current polarity via structural configuration, revealing the tunability of current flow in non-Hermitian systems. The maximum magnitude of the induced magnetic field reaches approximately $2.16$~T for $N=16$ and $1.92$~T for $N=26$, both corresponding to \textit{case~1}, underscoring the potential for high local magnetic response in $\mathcal{PT}$-symmetric architectures.

\begin{figure}[!htbp]
\centering
\includegraphics[width=8.8cm,height=8.0cm]{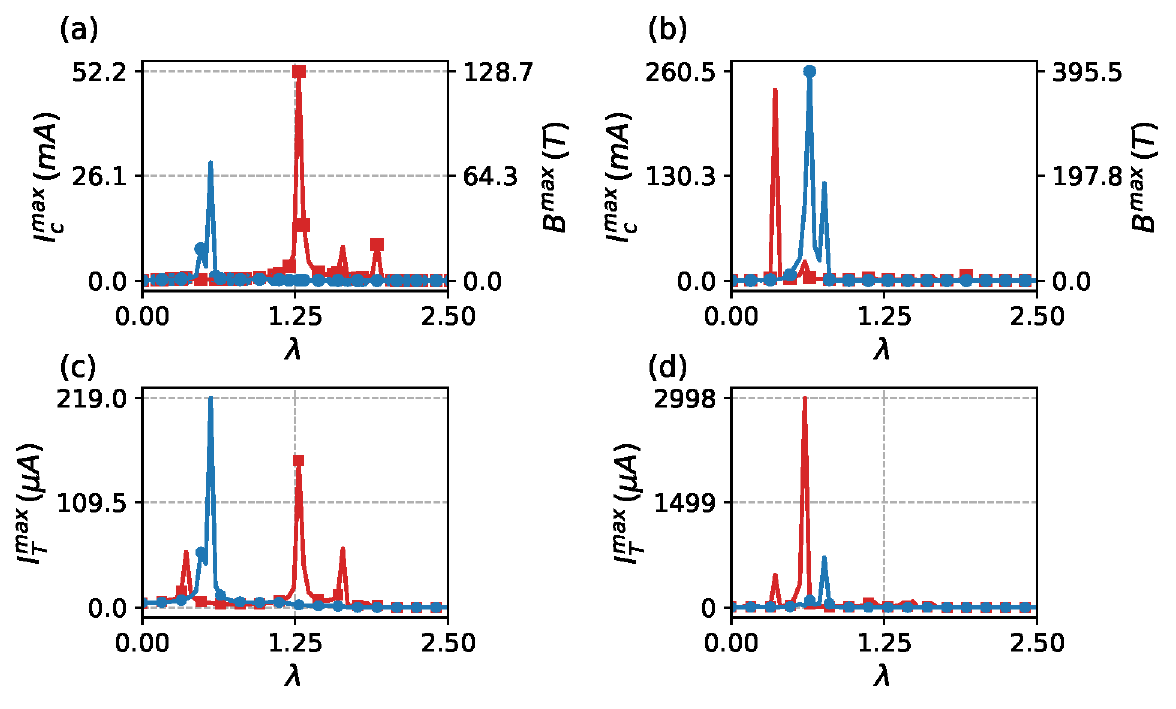}
\caption{(Color online).
Dependence of the maximum circular current and junction current on the gain-loss strength $\lambda$ in $\mathcal{PT}$-symmetric Fibonacci rings of two representative sizes ($N=16$ and $N=26$).
Panels~(a) and~(b) depict the evolution of the maximum circular current as $\lambda$ varies, while panels~(c) and~(d) show the corresponding junction current characteristics.
Red and blue curves correspond to case~1 and case~2 configurations, respectively, as defined in the main text.
The induced magnetic field, generated by the circulating current and shown on the twin-$y$ axes of panels~(a) and~(b).}
\label{fig:PT_NH_ICITMAX}
\end{figure}
\begin{figure*}[!htbp]
\centering
\includegraphics[width=0.98\textwidth]{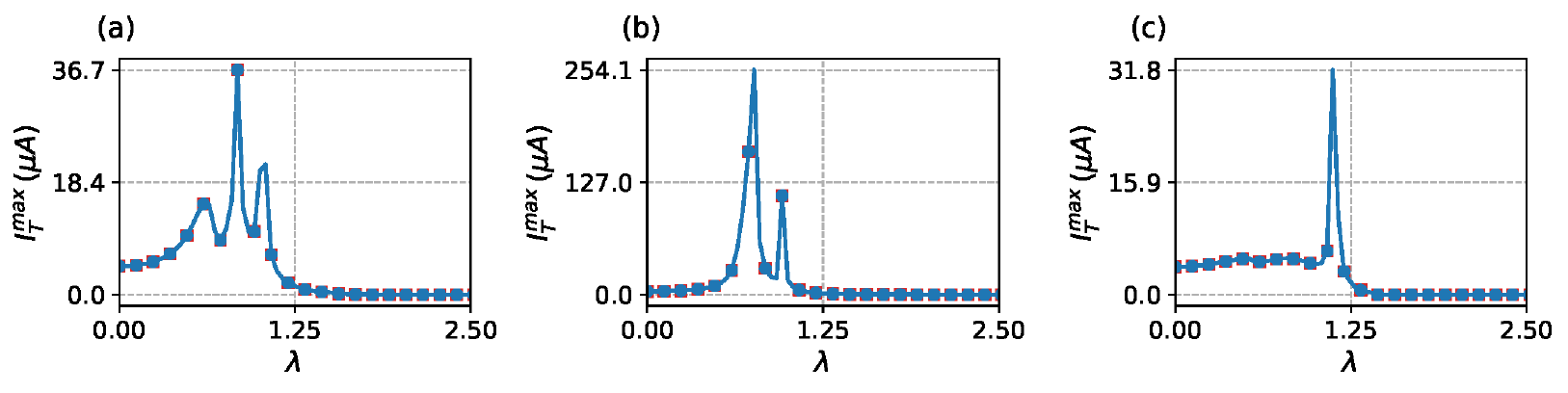}

\vspace{-3mm}

\includegraphics[width=0.98\textwidth]{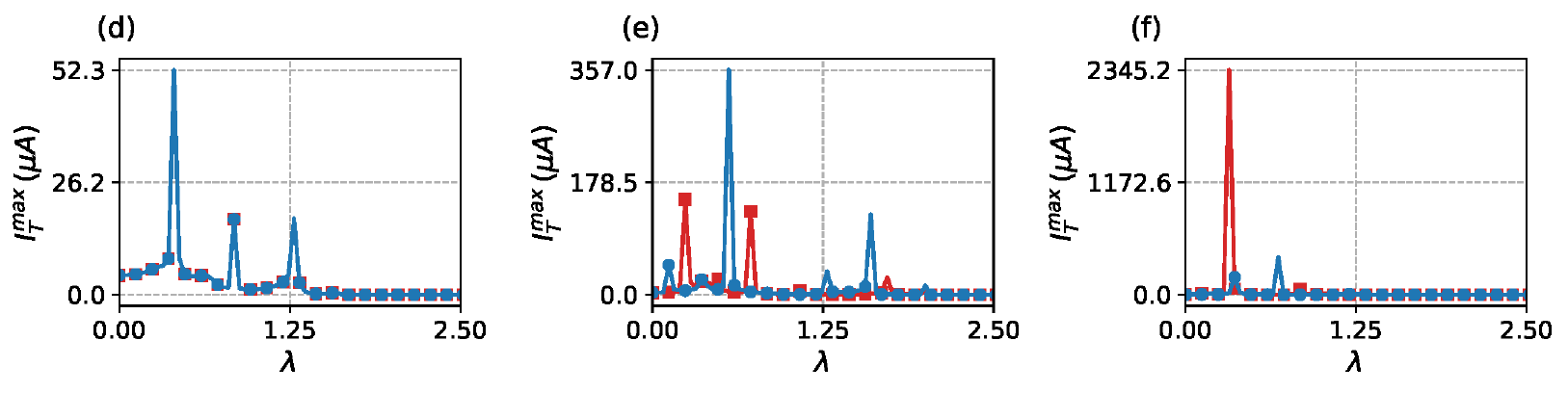}

\caption{(Color online). The variation of the maximum transport current, optimized over the applied bias window, is shown as a function of the gain–loss strength $\lambda$ for the non-$\mathcal{PT}$-symmetric configuration. The three columns correspond, from left to right, to the hopping regimes $t_1=t_2$, $t_1>t_2$, and $t_1<t_2$, highlighting the role of hopping asymmetry on non-Hermitian transport. The upper and lower panels represent system sizes $L=16$ and $L=26$, respectively, allowing a direct comparison of finite-size effects. Results for the two complementary realizations of on-site potential modulation (case~1 and case~2) are depicted by red and blue curves, respectively.}
\label{fig:nptitall}
\end{figure*}
To get overall response, we present a systematic analysis of how the maximum circular and junction currents evolve with the gain-loss strength $\lambda$ in the $\mathcal{PT}$-symmetric non-Hermitian Fibonacci rings in Figs.~\ref{fig:PT_NH_ICITMAX}(a)–(d). For each value of $\lambda$, the maximum current is obtained as the absolute peak current over the entire applied voltage range, allowing us to capture the most significant transport response of the system. The red and blue curves correspond to \textit{case~1} and \textit{case~2} configurations, respectively, while the left and right columns illustrate results for system sizes $N=16$ and $N=26$. A pronounced enhancement of both circular and junction currents is observed, leading to substantial magnetic field generation, highlighting the amplification effects induced by balanced gain and loss. For $N=16$, the circular current (and associated magnetic field) in \textit{case~1} surpasses that of \textit{case~2} across most of the $\lambda$ range, whereas the junction current displays the opposite trend, with \textit{case~2} yielding higher transport efficiency. Interestingly, for $N=26$, this behavior reverses, the circular current becomes more dominant in \textit{case~2}, while \textit{case~1} exhibits stronger junction current. This size-dependent inversion emphasizes the intricate competition between non-Hermitian amplification, Fibonacci modulation, and quantum interference pathways, collectively governing the transport and magnetic responses of $\mathcal{PT}$-symmetric Fibonacci systems. Moreover, the overall current magnitude in the non-Hermitian system is significantly enhanced compared to its Hermitian counterpart, owing to the gain–loss-induced amplification of transport channels.

Up to this point, we have restricted our analysis to the $\mathcal{PT}$ symmetric case and uniform-hopping limit $t_{1}=t_{2}$, which is already sufficient to produce a clear distinction between the transport responses of the two gain--loss configurations (\textit{cases $1$ \textnormal{and} $2$}), as evidenced by the preceding results. However, this simplification does not exhaust the full parameter space of the model, particularly in the non-$\mathcal{PT}$-symmetric regime where additional symmetry breaking mechanisms may become relevant. In the following subsection, we therefore extend the discussion to include both equal and unequal hopping scenarios, examining systems with $t_{1}=t_{2}$ as well as explicitly dimerized structures with $t_{1}\neq t_{2}$. This comparative analysis reveals that, while uniform hopping can distinguish the two cases in certain regimes, it is the introduction of hopping asymmetry that becomes essential for rendering \textit{cases} \textit{$1$} and \textit{$2$} indistinguishable, an outcome that cannot be achieved by the condition $t_{1}=t_{2}$ alone. The detailed numerical results supporting this statement are presented and analyzed in the subsequent subsection.

\subsubsection{Non $\mathcal{PT}$-symmetric configuration}
%subsubsection{Non Hermitian Case: Maximum Transport and Circular Current with $\lambda$}
\begin{figure*}[!htbp]
\centering
\includegraphics[width=0.98\textwidth]{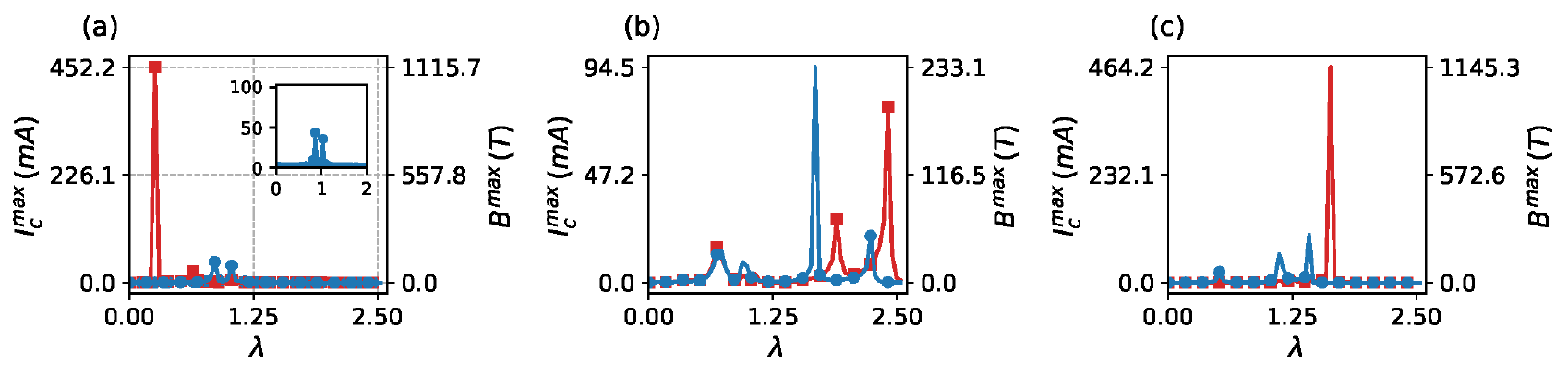}

\vspace{-3mm}

\includegraphics[width=0.98\textwidth]{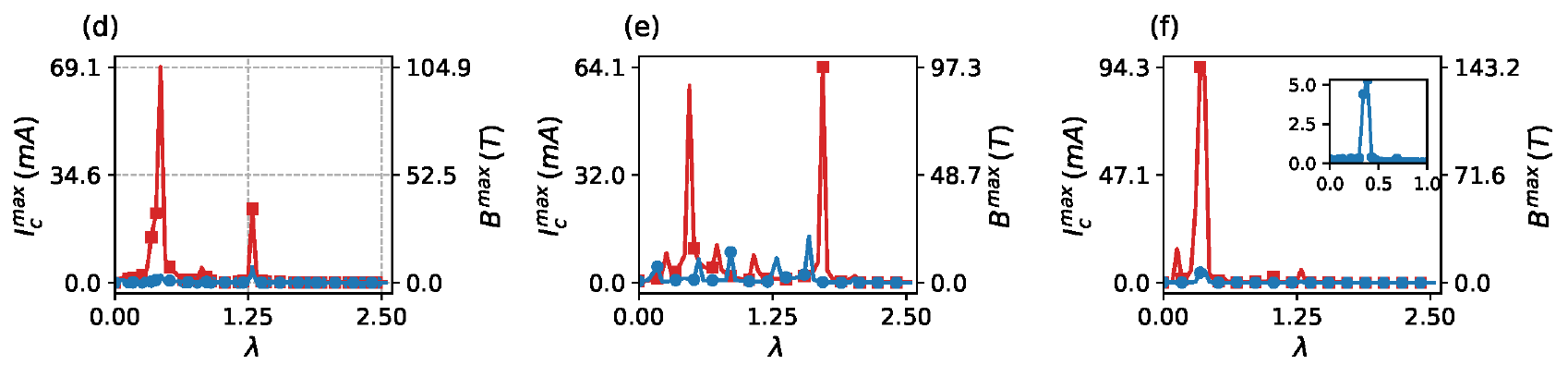}

\caption{(Color online). The dependence of the maximum circular current and the corresponding induced magnetic field, displayed on a twin $y$-axes, is presented as a function of the gain–loss strength $\lambda$ for the non-$\mathcal{PT}$-symmetric configuration. The column-wise arrangement follows the same hopping hierarchies as in the previous figures, namely $t_1=t_2$, $t_1>t_2$, and $t_1<t_2$, while the row-wise layout distinguishes between different system sizes, consistent with earlier plots.}
\label{fig:npticall}
\end{figure*}
In Figs.~\ref{fig:nptitall}(a)–(c), we illustrate the dependence of the maximum transport current on the gain–loss strength $\lambda$ for a finite system of size $L=16$, where the maximum value is obtained by scanning the full applied voltage window, while the corresponding results for a larger lattice with $L=26$ are shown in panels (d)–(f). The red and blue curves denote two distinct non-$\mathcal{PT}$-symmetric configurations, labeled as \textit{case~1} and \textit{case~2}, respectively, and the columns from left to right correspond to uniform hopping ($t_1=t_2$), forward dimerization ($t_1>t_2$), and reverse dimerization ($t_1<t_2$). A striking feature emerges for the smaller system size $L=16$, where the transport currents for \textit{case~1} and \textit{case~2} remain identical across the entire range of $\lambda$, independent of the underlying hopping correlations, indicating that finite-size coherence dominates over structural asymmetry in this regime. Upon increasing the system size to $L=26$, the same equivalence between the two cases persists only for the uniform hopping configuration $t_1=t_2$, whereas a clear separation between the two currents develops once hopping dimerization is introduced, i.e., for $t_1>t_2$ and $t_1<t_2$. This behavior highlights that, in non-$\mathcal{PT}$-symmetric systems, the manifestation of current asymmetry is not universal but instead emerges from a subtle interplay between system size and hopping correlations. In sharp contrast, for $\mathcal{PT}$-symmetric configurations, the currents corresponding to the two cases remain distinct for both system sizes and even in the absence of hopping dimerization ($t_1=t_2$), underscoring the fundamentally different role played by symmetry protection. These observations establish hopping dimerization as an essential ingredient for activating nontrivial transport asymmetry in non-$\mathcal{PT}$-symmetric setups, where spatial correlations become increasingly relevant.

It is worth emphasizing that the observed contrast between \textit{cases \textit{1}} and \textit{2} is also strongly influenced by the system size through the underlying Fibonacci geometry of the ring. For a ring with total size $L = 16$, corresponding to the Fibonacci generation comprising $8$ sites, both the upper and lower arms of the ring consist of an even number of lattice sites. In this configuration, the site at which the source is attached, whether it hosts gain or loss, is flanked on both sides by sites of the same character, i.e., gain (or loss), leading to a locally homogeneous non-Hermitian environment at the contact. An analogous situation arises at the drain end, where the loss (or gain) site is again surrounded by sites of identical nature. In sharp contrast, for a larger ring with $L=26$, corresponding to an odd Fibonacci generation ($13$ sites per arm), the local environment at the contacts becomes intrinsically asymmetric: the gain (or loss) site at the source is neighbored by one gain and one loss site, while the loss (or gain) site at the drain is similarly flanked by sites of opposite character. This mismatch in the local gain–loss landscape at the contacts, when combined with the correlated hopping structure imposed by the Fibonacci sequence, plays a decisive role in shaping the distinct transport signatures associated with \textit{cases} $1$ and $2$. Notably, the distinct signatures associated with even and odd numbers of sites in each arm of the Fibonacci ring persist for larger system sizes as well, a feature that will be systematically examined in the later part of this article.

%  \begin{figure*}[!htbp]
% \centering
% \includegraphics[width=18.5cm]{NOTPT_IT_SIZE_ALL.eps}
% \caption{(Color online) Size-dependent transport properties for the non-Hermitian case.}
% \label{fig:size_combined}
% \end{figure*}

\begin{figure*}[!htbp]
\centering
\includegraphics[width=0.98\textwidth]{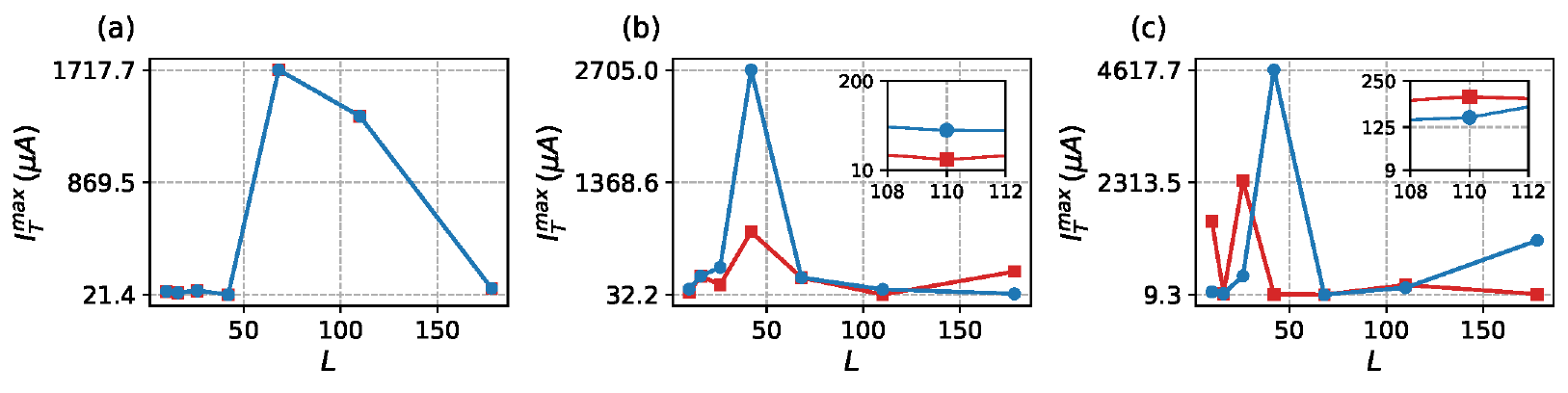}

\vspace{-3mm}

\includegraphics[width=0.98\textwidth]{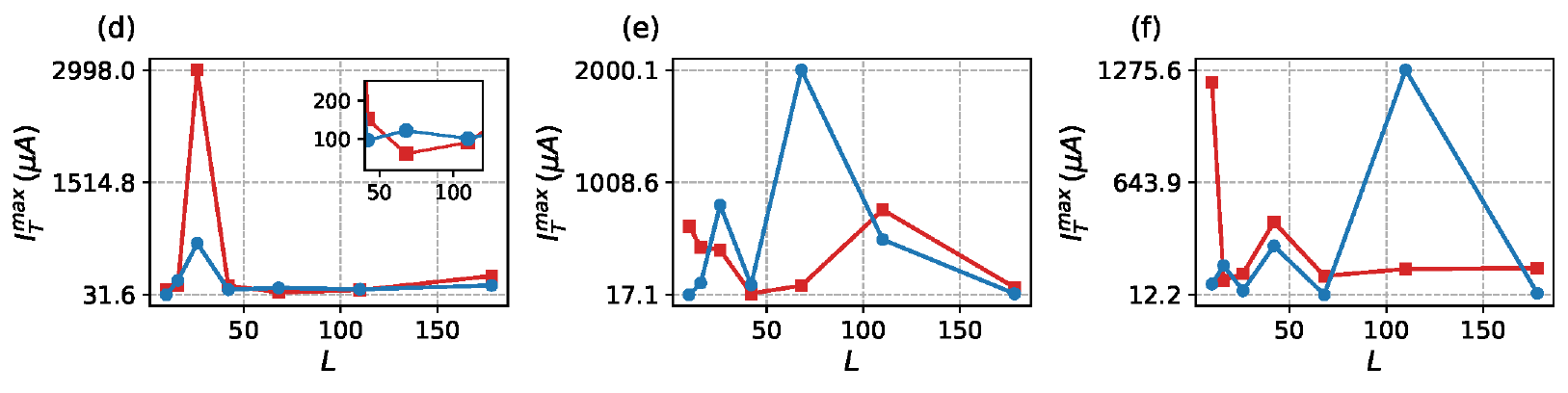}

\caption{(Color online). The system-size dependence of the maximum transport current is shown for both non-$\mathcal{PT}$-symmetric (upper panel) and $\mathcal{PT}$-symmetric (lower panel) configurations. The columns, from left to right, correspond to the three distinct hopping scenarios, namely $t_1=t_2$, $t_1>t_2$, and $t_1<t_2$. Red and blue curves represent the two complementary cases considered in this work. For each system size $L$, the reported current corresponds to the global maximum obtained by scanning over the gain–loss strength $\lambda$; furthermore, for every fixed $\lambda$, the current itself is evaluated as the maximum over the entire applied voltage window.}
\label{fig:SIZEIT}
\end{figure*}

% \begin{figure*}[!htbp]
% \centering
% \includegraphics[width=0.98\textwidth]{NOTPT_IC_SIZE_ALL.eps}

% \vspace{-3mm}

% \includegraphics[width=0.98\textwidth]{PT_IC_SIZE_ALL.eps}

% \caption{(Color online). The figure illustrates how the circular current and the concomitant induced magnetic field evolve with increasing system size for both non-$\mathcal{PT}$-symmetric (upper panel) and $\mathcal{PT}$-symmetric (lower panel) realizations. The three columns correspond to different hopping correlations, namely the uniform ($t_1=t_2$), forward-dimerized ($t_1>t_2$), and backward-dimerized ($t_1<t_2$) limits. Results for the two distinct potential configurations are distinguished by red and blue curves.}
% \label{fig:SIZEIC}
% \end{figure*}
  
%$\bullet$\emph{{Transmission characteristics and junction current:}}
% \begin{figure*}[!htbp]
% \centering
% \includegraphics[width=14cm]{JEICV_H.pdf}
% \caption{(Color online) Transmission and current characteristics for the Hermitian case.}
% \label{fig:nptjeicvh}
% \end{figure*}
% \begin{figure*}[!htbp]
% \centering
% \includegraphics[width=0.98\textwidth]{16ICMAXB_HERM.pdf}

% \vspace{-3mm}

% \includegraphics[width=0.98\textwidth]{26ICMAXB_HERM.pdf}

% \caption{Hermitian case: Circular current for different ring sizes.}
% \label{fig:npticmaxherm}
% \end{figure*}
In Figs.~\ref{fig:npticall}(a)–(c), we present the variation of the maximum circular current as a function of the gain/loss strength $\lambda$ for a system size $L=16$, where the maximum is extracted over the entire applied voltage window, while panels (d)–(f) display the corresponding results for a larger system with $L=26$. The red and blue curves represent two distinct non-$\mathcal{PT}$-symmetric configurations, referred to as \textit{case~1} and \textit{case~2}, respectively. From left to right, the columns correspond to uniform hopping ($t_1=t_2$), dimerization favoring stronger intra-cell hopping ($t_1>t_2$), and the reverse dimerization regime ($t_1<t_2$). The magnetic fields generated by the circulating currents, evaluated self-consistently from the current amplitudes, are shown on the secondary $y$-axis for direct comparison. For both system sizes and for both cases, the maximum circular current exhibits a pronounced nonmonotonic dependence on $\lambda$, characterized by alternating peaks and dips that signal the delicate competition between gain-loss modulation and coherent transport. Notably, while finite circular currents persist in this non-$\mathcal{PT}$-symmetric setup, their overall magnitudes are systematically increased and reduced compared to the corresponding $\mathcal{PT}$-symmetric configurations for $L=16$ and $L=26$ respectively, reflecting enhanced backscattering and reduced constructive interference. Nevertheless, the distinction between \textit{case~1} and \textit{case~2} remains markedly amplified relative to their Hermitian counterparts (verified separately and not shown here), underscoring the pivotal role of non-Hermiticity in selectively enhancing current asymmetry even in the absence of exact $\mathcal{PT}$ symmetry. An additional noteworthy outcome concerns the behavior of the circular current, which, in contrast to the transport current, does not exhibit any pronounced sensitivity to whether the number of sites in each arm of the ring is odd or even. In this case, the condition $t_1=t_2$ by itself is sufficient to yield a clear and robust distinction between the two complementary gain--loss configurations (\textit{case~1} and \textit{case~2}), irrespective of the underlying parity of the Fibonacci sequence length. This insensitivity to arm parity highlights the fundamentally different mechanisms governing circulating and transport currents, and underscores that, for circular current generation, uniform hopping already provides adequate asymmetry when combined with non-Hermitian gain--loss engineering.

\subsubsection{System-size dependence of the transport current for both $\mathcal{PT}$-symmetric and non-$\mathcal{PT}$-symmetric configurations}

To assess the robustness of the transport-current signatures associated with \textit{case~1} and \textit{case~2}, particularly their sensitivity to the parity (odd or even) of the number of sites in each arm of the ring, we perform a systematic size-scaling analysis of the junction current in non-Hermitian Fibonacci rings. The analysis covers several Fibonacci generations and increasing system sizes, with the corresponding results displayed in Figs.~\ref{fig:SIZEIT}(a)–(f) for non-$\mathcal{PT}$-symmetric (upper panels) and $\mathcal{PT}$-symmetric (lower panels) configurations. Such a comparative framework allows us to critically examine the persistence and stability of these parity-driven features against increasing system size and enhanced gain--loss complexity. The ring size is constructed following successive generations of the Fibonacci sequence, with the upper and lower arms of the ring each obeying the same quasiperiodic order; for instance, a Fibonacci index of $13$ corresponds to $13$ sites in each arm, leading to a total system size of $L=26$, and this construction is systematically extended up to the Fibonacci number $89$ ($L=178$). For a given lattice size and hopping configuration, the transport current is evaluated over the full bias voltage window to identify its maximal value, which is subsequently optimized over the entire range of the gain–loss strength $\lambda$ to extract the global peak current. This procedure ensures that the reported current faithfully represents the most efficient transport regime accessible to the system and allows a transparent comparison of how Fibonacci pattern, non-Hermiticity, and symmetry constraints collectively govern the evolution of the junction current with increasing system size. Now in Figs.~\ref{fig:SIZEIT}(a)–(c), we examine the evolution of the transport current as a function of the system size $L$ for non-$\mathcal{PT}$-symmetric scenario at three representative hopping correlations discussed earlier, with the red and blue curves corresponding to \textit{case~1} and \textit{case~2}, respectively. Several nontrivial trends emerge from this size-scaling analysis. For uniform hopping, $t_1=t_2$ [subplot (a)], the transport current remains identical for both cases over the entire range of system sizes, indicating that in the absence of hopping asymmetry the underlying structure alone is insufficient to distinguish between the two cases. The situation changes qualitatively once hopping dimerization is introduced, i.e., for $t_1\neq t_2$, where the current response becomes highly sensitive to the parity of the Fibonacci generation. In particular, the currents corresponding to \textit{case~1} and \textit{case~2} coincide when the Fibonacci index is even (e.g., $8, 34, \ldots$), which translates into total system sizes such as $L=16, 68, \ldots$, whereas a clear separation between the two currents develops for odd Fibonacci indices (e.g., $5, 13, 21, 55, 89, \ldots$), corresponding to system sizes $L=10, 26, 42, 110, 178, \ldots$. This parity-dependent distinction is most pronounced in the regime $t_1<t_2$ [subplot (c)], although a similar, albeit weaker, behavior persists for $t_1>t_2$ [subplot (b)]. Notably, in the latter case the currents for certain sizes, such as Fibonacci number $55$ (or $L=110$), may appear nearly overlapping at first glance; however, a closer inspection, provided through the inset, reveals a finite but subtle difference. These observations collectively demonstrate that the interplay between hopping correlations and the even–odd character of the Fibonacci sequence plays a decisive role in determining whether the transport currents for the two cases remain identical or become distinct, thereby highlighting an unconventional size-parity effect intrinsic to quasiperiodic lattices.

In contrast to the non-$\mathcal{PT}$-symmetric scenario, the $\mathcal{PT}$-symmetric configuration, illustrated in Figs.~\ref{fig:SIZEIT}(d)–(f), does not exhibit any comparable sensitivity of the transport current to the even–odd character of the underlying Fibonacci sequence. For all three hopping correlations considered, the currents associated with \textit{case~1} and \textit{case~2} evolve with system size and remain distinct, without displaying the alternating pattern tied to the parity of the Fibonacci generation observed earlier. This absence of Fibonacci-parity dependence reflects the constraining role of $\mathcal{PT}$ symmetry, which enforces a balanced distribution of gain and loss and preserves a robust correspondence between the two configurations, effectively suppressing size-induced interference effects that would otherwise differentiate them. As a result, the transport response in the $\mathcal{PT}$-symmetric regime is governed predominantly by symmetry protection rather than by subtle Fibonacci correlations, highlighting a fundamental qualitative distinction between $\mathcal{PT}$-symmetric and non-$\mathcal{PT}$-symmetric transport in Fibonacci lattices.

\section{Conclusion}

In summary, we have presented a comprehensive theoretical investigation of transport and circulating currents, along with the associated induced magnetic fields, in Hermitian and non-Hermitian Fibonacci quantum rings, where the non-Hermitian regime is further realized in both $\mathcal{PT}$-symmetric and non-$\mathcal{PT}$-symmetric configurations. The system is engineered by assigning real on-site potentials $\pm \lambda$ in the Hermitian limit and by introducing physically balanced gain and loss, $\pm \mathrm{i}\lambda$, arranged according to a Fibonacci sequence in the non-Hermitian case. By judiciously distributing these non-Hermitian elements between the two arms of the ring, we achieve controlled preservation or explicit breaking of $\mathcal{PT}$ symmetry. Further tunability is achieved by interchanging the signs of the on-site potentials, enabling a detailed assessment of how configurational asymmetry, aperiodicity, and hopping correlations jointly govern transport characteristics. Within the NEGF framework, we evaluate transmission spectra, junction currents, and bond-resolved circulating currents, the latter giving rise to an effective magnetic field threading the ring. We begin with the Hermitian limit as a natural reference point and progressively extend the analysis to the non-Hermitian regime in order to elucidate the decisive role played by engineered gain--loss patterns in shaping the transport response of the system.
%By benchmarking against the Hermitian limit, we explicitly demonstrate that a perfectly ordered and symmetrically contacted ring fails to sustain any finite circular current, and even moderate disorder induces only marginal magnetic response. In striking contrast, the non-Hermitian regime exhibits a pronounced enhancement of both transport and loop currents, underscoring the pivotal role of gain–loss engineering in activating otherwise forbidden current circulation. Our results further reveal that $\mathcal{PT}$ symmetry generally promotes stronger current amplification and more robust magnetic response, while symmetry breaking introduces heightened sensitivity to system size, Fibonacci parity, and hopping asymmetry. Collectively, these findings establish non-Hermiticity, when combined with quasiperiodic topology, as a powerful and versatile mechanism for tailoring current flow and magnetic functionality in symmetric mesoscopic ring geometries, opening new avenues for designing active quantum devices with controllable current circulation and field generation. 
The principal conclusions of this work can be articulated as follows.

$\bullet$ In the Hermitian regime, only a marginal enhancement of both the transport current and the circular current is observed upon increasing the potential strength $\lambda$. Although the circular current exhibits weak resonant-like peaks as a function of $\lambda$, these features are not sustained and the current ultimately diminishes for larger potential strengths. In contrast, the transport current shows a monotonic suppression from the outset with increasing $\lambda$, reflecting the dominant role of enhanced backscattering and localization induced by the growing quasiperiodic potential. Overall, the Hermitian system fails to support robust current amplification at symmetrical lead--ring--lead connection, underscoring the limited efficacy of purely real potentials in sustaining transport and circulating currents.

$\bullet$ We demonstrate that the introduction of non-Hermiticity through balanced gain and loss leads to a substantial amplification of the transport current, circular current, and the corresponding induced magnetic field when compared to the Hermitian limit. This enhancement underscores the effectiveness of non-Hermitian engineering as an active control knob for tailoring quantum transport, enabling current magnitudes that are otherwise unattainable in conventional passive systems.

$\bullet$ A comparative analysis of the two symmetry classes reveals that both the $\mathcal{PT}$-symmetric and non-$\mathcal{PT}$-symmetric configurations can support appreciable current responses. While the $\mathcal{PT}$-symmetric setup exhibits enhanced transport in certain parameter regimes, there also exist regimes where the non-$\mathcal{PT}$-symmetric configuration yields a larger current. This interplay highlights the parameter dependent role in shaping coherent transport characteristics in open quantum systems.

$\bullet$ We further find that reversing the sign of the on-site potentials induces a markedly stronger modification of transport characteristics in the non-Hermitian regime than in Hermitian systems, reflecting an enhanced sensitivity to local potential rearrangements that originates from the complex energy spectrum and the redistribution of probability amplitudes inherent to non-Hermitian dynamics.

$\bullet$ In the absence of $\mathcal{PT}$ symmetry, the transport response exhibits a pronounced dependence on system size, which can be traced back to the parity (odd or even) of the underlying Fibonacci sequence and is further modulated by correlated hopping processes. This observation establishes a direct link between Fibonacci ordering, lattice parity, and non-Hermitian transport behavior.

$\bullet$ Finally, in sharp contrast to the widely reported monotonic decay of current with increasing system size in Hermitian quasiperiodic or disordered rings, the non-Hermitian Fibonacci rings investigated here display a distinctly nonmonotonic size dependence. This unconventional scaling behavior reveals the intricate interplay among aperiodicity, topology, and non-Hermiticity, and points toward new avenues for designing mesoscopic and nanoscale devices with tunable and size-resilient transport characteristics.

\end{document}